\documentclass[aip,prm,twocolumn,superscriptaddress,floatfix,10pt]{revtex4}
\usepackage{graphicx}
\usepackage{amsmath}
\usepackage[bbgreekl]{mathbbol}
\usepackage{color}
\usepackage{soul}
\usepackage{ulem}
\begin{document}

\title{Computing finite--temperature elastic constants with noise cancellation}

\author{Debashish Mukherji}
\affiliation{Institut f\"ur Theoretische Physik, Georg--August--Universit\"at G\"ottingen, 37077 G\"ottingen, Germany}
\author{Marcus M\"uller}
\affiliation{Institut f\"ur Theoretische Physik, Georg--August--Universit\"at G\"ottingen, 37077 G\"ottingen, Germany}
\author{Martin H. M\"user}
\affiliation{Dept. of Materials Science and Engineering, Saarland University, 66123 Saarbr\"ucken, Germany}


\begin{abstract}
Elastic constants are central material properties, frequently reported in experimental and theoretical studies.
While their computation is straightforward in the absence of thermal fluctuations, finite--temperature 
methods often suffer from poor signal--to--noise ratios, the presence of strong anharmonic effects, 
or require second-order derivatives with respect to spatial coordinates to be taken.
Here, we show how to compute elastic constants in thermal ordered and disordered systems by generalizing a
noise--cancellation method originally developed for piezoelectric coupling coefficients.
A slight strain is applied to an equilibrated solid.
Simulations of both the strained and unstrained (or oppositely strained) reference systems are performed using identical thermostatting schemes.
As demonstrated theoretically and with generic one--dimensional models, 
this allows stress differences to be evaluated and elastic constants to be determined with 
favorable thermal noise.
We then apply this approach across a diverse set of systems, spanning crystalline argon, ordered silicon as well 
as amorphous silicon, poly(methyl methacrylate), and cellulose derivatives.
\end{abstract}

\maketitle

\section{Introduction}

Elastic constants are routinely computed in molecular simulations using two mainstream methods:  
response--to--deformation (RTD)~\cite{Born1954book} and fluctuation--based (FB)~\cite{Parrinello82jcp,Ray1984jcp} computations.  
Their central ideas are as follows:  
When deforming a solid from a stress--free state, the strain tensor elements, $\varepsilon_i$ in Voigt notation, 
are known, and the resulting stress--tensor elements, $\sigma_i$, can be computed. 
The elastic tensor elements, $C_{ij}$, then follow from $\sigma_i = C_{ij} \varepsilon_j$. 
Alternatively, the simulation box can be fixed, and an estimator be evaluated, which consists of a so-called Born term and a stress-fluctuation correction~\cite{Squire1969P,Lutsko1989JAP}.
The Born term hinges on the Hessian matrix and is thus frequently not available, in particular for many-body or machine-learned potentials.
%
%
Analyzing box--shape fluctuations in constant--stress simulations is also a viable approach~\cite{Parrinello82jcp}
provided that meaningful barostats are employed, which can be difficult to find in standard software. 
Worth mentioning is also the determination of elastic constants through the computation of the sound velocity~\cite{Huang1950PRSL,Barron1965PPS}, 
which can be regarded as a FB method.
This approach, however, relies on analyzing dynamical rather than static covariances. 

Each of the mainstream methods includes further sub--methods~\cite{Shinoda04PRB,Tadmor2011Book}.  
For example, at zero temperature it suffices to compute the energy as a function of strain, although internal modes 
must be allowed to relax in a non--affine fashion if applicable~\cite{Born1954book,Tadmor2011Book}.    
Alternatively, one can compute second--order derivatives of the energy with respect to atomic coordinates, where non--affine 
displacements can again cause complications~\cite{Born1954book,Tadmor2011Book}, in particular for a stressed reference~\cite{Griesser23prm}.
Additional modifications apply in the presence of thermal fluctuations~\cite{Squire1969P,Ray1988CPR,Lutsko1989JAP} or, again, when the reference is stressed~\cite{Griesser23prm}.
Accounting for quantum fluctuations, not considered here, requires additional terms~\cite{Schoffel2001PRB,Folkner2024JAP}.

RTD and FB computations of elastic tensor elements each have their own advantages and disadvantages.  
In particular, FB computations require long convergence times; 
using larger system sizes even increases the stochastic error, 
given fixed compute time, while the RTD method would leave it constant.  
However, RTD computations using simple forward-difference schemes and conducted at what may deceptively be considered a small strain, say, 1\%, 
can produce errors in the elastic constants on the order of 10\% due to nonlinear effects, whereas simulations at smaller strains are dominated by stochastic noise.

An alternative to the above--mentioned algorithms is based on the diagonalization of the Hessian matrix while accounting for non--affine deformations in the system. 
Although such approaches are computationally expensive, particularly for large system sizes, they yield exact results 
in the absence of thermal fluctuations and allow local elastic constants to be defined~\cite{Maloney2006PRE, Tsamados2009PRE}.
When the temperature is finite, anharmonic effects set in, which can be captured by considering the Hessian of the 
instantaneous rather than the relaxed normal modes~\cite{Vaibhav2024M}.
However, the occurrence of negative eigenvalues in the Hessian 
makes a rigorous error estimate difficult.
Even the computation of viscoelastic properties of ultra-strong disordered systems, which lack barrier crossings, is complex when done with care~\cite{Griesser2024PRE}. 
In this context, the present work focuses on a lean computation of only the elastic tensor, prioritizing minimal implementation, low noise, and minimal systematic errors.

Here, we extend a noise--cancellation technique originally developed for the piezoelectric coefficient $\gamma$ of $\alpha$--quartz, 
which remained effective even near the $\alpha$--$\beta$ quartz transition temperature~\cite{Herzbach06cpc}.
The method contrasted the strain in two constant--stress simulations, both initialized from the same equilibrated 
configuration and both using the same random number sequence in the thermostat.
In one sample, an electric field was applied, which slowly increased over time, while the other sample remained field--free.
The strain difference between the samples was essentially free of thermal fluctuations-- at least during the initial stages 
of the simulation-- and allowed $\gamma$ to be deduced reliably within a few hundred time steps.
This approach can be classified as a correlated sampling technique, which is more commonly employed in 
Monte Carlo than in molecular dynamics simulations, and was potentially first formulated in the context of nuclear physics~\cite{Rief1984ANE}.

For the computation of elastic constants, we impose a minuscule constant strain rather than a slowly ramped field, as in the original work.
This simplifies the procedure, since it avoids the complications of maintaining constant stress.
Two simulations are performed from the same equilibrated configuration and using identical random number sequences in the thermostat.
The strain difference between the simulations is essentially free of thermal fluctuations-- 
at least during the initial stages; however, care must be taken to suppress the effects of decorrelation at longer times.

The remainder of this paper is organized as follows:
Section~\ref{sec:ThMe} introduces the fundamental concepts underlying the pursued method,
while the relevant material--specific details are given in Section~\ref{sec:method}.
Section~\ref{sec:res} presents the results.
Conclusions are drawn in Section~\ref{sec:conclusions}.

\section{Theory of noise cancellation} 
\label{sec:ThMe}

\subsection{Preliminary considerations}

Before elaborating on the ideas behind noise cancellation, it is worth substantiating our claim from the introduction that 
a 1\% strain can change the elastic constants by more than 10\%.
As a simple illustration, we consider the Lennard--Jones (LJ) potential in its $m-n$ form~\cite{Muser2022APX}, ($m=12$, $n=6$), 
\begin{equation}
U(r) = U_0 \left[ \left( \frac{r_0}{r}\right)^{12} - 2 \left( \frac{r_0}{r}\right)^{6}\right],
\end{equation}
where $r$ is the distance between two atoms, $U_0$ the potential depth, and $r_0$ the equilibrium distance.
The corresponding bond stiffness is
\begin{equation}
k_{12,6}(r) = \frac{12 U_0}{r_0^2}\left[ 13 \left( \frac{r_0}{r}\right)^{14} - 7 \left( \frac{r_0}{r}\right)^{8} \right].
\end{equation}
Using standard relations to connect such bond stiffnesses to elastic--tensor elements in simple solid 
structures~\cite{Muser2022APX}, one finds that the ratio $k_{12,6}(0.99r_0)/k_{12,6}(r_0) \approx 1.23$ corresponds to roughly a 20\% increase in the bulk modulus $B$.

For covalently bonded systems, the stiffening effect of the same compressive strain (1\%) is smaller.
For example, for the regular 2--1 (nearest--neighbor) Morse potential,
\begin{equation}
k_{2,1}(r) = \frac{2 U_0}{r_0^2}\left[ 3 \left( \frac{r_0}{r}\right)^{4} - \left( \frac{r_0}{r}\right)^{3} \right],
\end{equation}
one obtains $k_{2,1}(0.99r_0)/k_{2,1}(r_0) \approx 1.046$.
Even so, an error approaching 5\% can still be considered unsatisfactory.

The preliminary considerations discussed above highlight why computing the elastic constants of (amorphous) polymers presents 
particular challenges. Specifically, the steepness of the Lennard--Jones (LJ)--like interactions necessitates the use of very 
small strain increments in finite--difference (FD)--based schemes for reliable response--to--deformation (RTD) calculations. 
At the same time, the relatively low energy scales-- characterized by small interaction prefactors 
(e.g., $U_0$ typically being below the thermal energy at room temperature)-- make the system highly susceptible to thermal fluctuations. 
These combined effects contribute to significant noise in stress responses and complicate the accurate determination of the elastic tensor components.


\subsection{Noise cancellation in ideally harmonic systems}
\label{sss:noinse_harm}

To discuss the generic features of noise cancellation in harmonic and, to some degree, near--harmonic (extended) systems, 
we focus on a single but periodically repeated diatomic molecule in a simulation box.
This molecule has a rigid, short bond ($s$) and a soft, long bond ($l$), both of which are approximated as harmonic,
which captures the essential features of a linear, harmonic solid undergoing non--affine deformation under strain.
The setup can be seen as a minimalist description of a zero--wavenumber optical vibration in an extended one--dimensional linear chain of H$_2$ molecules.
A more realistic example would be the optical $\Gamma$--point vibration parallel to the direction of the 
Peierls distortion undergone by antimony from a simple cubic reference along [111].

The goal of the treatment presented hereafter is to understand how the mean stress and its standard deviation depend on time (shortly) after applying strain to an equilibrated sample.
This proves useful to rationalize results and to identify first guidelines for convenient thermostat and parameter choices.
For example, it will become clear why global kinetic energy controls (e.g., canonical sampling by velocity rescaling or No\'se--Hoover) 
might not benefit from noise cancellation as much as stochastic thermostats do.

Following the discussion in the preceding paragraph, we introduce a minimal Hamiltonian for the enthalpy per unit cell,
\begin{equation}
\label{eq:ideal-harm-hamil}
H = \frac{k_s}{2}\left(hx_{21}-a_s^0\right)^2 + \frac{k_l}{2}\left\{h(1-x_{21})-a_l^0\right\}^2 - h \sigma,
\end{equation}
where $k_s > k_l$ and $a_s^0 < a_l^0$ are the stiffnesses and equilibrium bond lengths of the short ($s$) and long ($l$) bonds, respectively.
The box length per unit cell is denoted by $h$, in analogy to the $h$-matrix of constant--stress simulations~\cite{Parrinello1981JAP}, and $x_{21} = x_2 - x_1$.
Here, $x_i$ are reduced coordinates, with real positions $X_i = h x_i$.
They do not appear in the final expressions but are required for intermediate steps.
Reduced coordinates are also the natural variables in the constant--stress ensemble, ensuring a direct link between the analytical treatment and simulations.
Finally, $\sigma$ is the external (tensile) stress.
For readability, the kinetic-energy
contribution is omitted here but will be included where necessary.

Minimizing $H$ with respect to $h$, while keeping $x_i$ (rather than $a_s$) constant, yields (the virial estimator for) the stress:
\begin{equation}
\label{eq:stress_estimator}
\sigma = - \frac{1}{h}\sum_{b=s,{l}}F_b a_b,
\end{equation}
the generalization of which to higher dimensions is commonly used to deduce stress--tensor elements in molecular simulations.
At a given value of $h$, a short--bond length of
\begin{eqnarray}
a_s^\text{eq}(h) = \frac{k_s}{K} a_s^0 + \frac{k_l}{K}(h-a_l^0)
\end{eqnarray}
minimizes the energy, where $K = k_s + k_l$.
Thus, the initial excess length of a short bond after an affine deformation is proportional to the applied strain $\varepsilon$, specifically
\begin{subequations}
\begin{eqnarray}
\Delta a^\text{eq}_s(t=0^+)  & \equiv & (1+\varepsilon)a_s{^{\text{eq}}}(h) - a_s{^{\text{eq}}}\{(1+\varepsilon) h\}  \\
& = & \underbrace{\frac{1}{K}(k_s a_s^0 - k_l a_l^0)}_{a_s'} \, \varepsilon,
\label{eqn:4b}
\end{eqnarray}
\end{subequations}
where the proportionality factor $a_s'$ is commonly positive, since bond--length ratios are of order unity, 
while the stiffnesses of short (e.g., covalent) and long (e.g., dispersive) bonds tend to differ substantially.

Here, and in the following, $\Delta$ refers to a difference between an observable in the strained and reference 
(unstrained or oppositely strained) sample, e.g., $\Delta f \equiv f(0<\varepsilon\ll 1)-f(0)$ denotes the difference 
between an observable at small finite strain and at zero strain.
The quantity $f$ is generally time dependent and can also be a compound property, such as products of forces and bond-lengths, $Fa$, to be considered further below. 
An exception to this rule is the time step $\Delta t$. 

Assume now that a \textit{thermal} solid, or an ensemble of equilibrated solids, is strained in an affine fashion 
from a reference shape $h^\text{ref}$ to $h^\text{str}=(1+\varepsilon) h^\text{ref}$ at the time $t = 0$.
Both the strained solid (str) and the unstrained reference (ref) will then be propagated in time 
using the same random process, e.g., one based on Brownian or Langevin dynamics.
To facilitate the treatment of the ensuing dynamics, we exploit the fact that our one--dimensional, 
two--body problem decomposes into the force--free center--of--mass motion and 
one internal eigenmode $a(t)$, e.g., $a_s(t)$, although $a_l(t)$ would have been an equally valid choice.
The internal eigenmode(s) then obey(s) an equation of motion having the form
of a harmonic oscillator with damping and noise,
\begin{equation}
\label{eq:eom_eigen_mode}
\mu \delta \ddot{a}(t) + \frac{\mu}{\tau} \delta\dot{a}(t) + K \delta a(t) = \Gamma(t),
\end{equation}
where $\delta a = a_s-a_s^{\rm eq}\{(1+\varepsilon)h\}$.
Although it is the analogues of the $x_n$ that are propagated in time by the MD code, we can deduce 
the resulting dynamics by scrutinizing Eq.~\eqref{eq:eom_eigen_mode}. 
In our specific example, $\mu$ is the reduced mass.
The friction coefficient, $\mu/\tau$, defines a the damping time scale, $\tau$.
The thermal force $\Gamma(t)$ must have a vanishing first moment and otherwise obey
$\langle \Gamma(t) \Gamma(t') \rangle = 2 \mu k_{\rm B} T \delta(t-t')/\tau$ to satisfy
the fluctuation--dissipation theorem.

Just before the strain is applied, i.e., at time $t = 0^-$, the bond variable $a_s(t)$ can 
be decomposed into the equilibrium value, $\langle a_s(t=0^-)\rangle = a_s^\text{eq}(h^\text{ref})$, and 
a thermal fluctuation $a_s(t=0^-) - a_s^\text{eq}(h^\text{ref})= \delta a^{\text{ref}-}$.
Once the strain is applied, the bond variable is given by,
\begin{subequations}
\begin{eqnarray}
a_s^\text{str}(t=0^+) & = & (1+\varepsilon) \left\{a^\text{eq}(h^\text{ref}) + \delta a^{\text{ref}-}\right\} \\
& = & a_s^\text{eq}(h^\text{str}) + \delta a^{\text{str}+}.
\end{eqnarray}
\end{subequations}
This defines the new, initial bond excess $\delta a^{\text{str}+}=(1+\epsilon) \delta a^{\text{ref}-} +(1+\epsilon) a_s^\text{eq}(h^\text{ref})- a_s^\text{eq}(h^\text{str})$. 

At times $t>0$, the formal solution of Eq.~\eqref{eq:eom_eigen_mode} reads
\begin{eqnarray}
\label{eq:formal_solution}
\delta a^\text{str}(t) = \delta a^{\text{str}+} C(t) + \underbrace{\int_0^t \mathrm{d}t' \, G(t-t') \Gamma(t')}_{\equiv g(t)},
\end{eqnarray}
where $C(t)$ and $G(t-t')$ are the relaxation and the Green's function of the damped harmonic oscillator, respectively. 
Note that the thermostat--noise term, $g(t)$, is independent of the starting condition.
Thus,  
\begin{equation}
\label{eq:formal_solution_2}
\Delta a_\text{MD}(t) = a^\text{eq}(h^\text{str}) - a^\text{eq}(h^\text{ref}) +
\left\{ \delta a^{\text{str}+} - \delta a^{\text{ref}-}  \right\} C(t).
\end{equation}
This is a deterministic function, except for the thermal noise present at $t = 0^-$, which vanishes on average.
Thus, the bond length difference becomes a noise--free observable in the long run (i.e., for $t\gg \tau$) but only if $C(t)$ decays to zero.
This is typically not the case in the microcanonical ensemble, corresponding to vanishing friction, $\tau \to \infty$, in Eq.~\eqref{eq:eom_eigen_mode}, 
or, when using a thermostat based on global kinetic energy control, reinforcing arguments that such schemes do not always function as thermostats.

The absence of ``new" ($t > 0$) random noise in $\Delta a_\text{MD}(t)$ might suggest a lack 
of noise in the stress estimator.
However, this is not the case. The virial stress estimator depends on products of bond lengths and bond 
forces, both of which evolve dynamically over time. Consequently, fluctuations persist in the stress estimator, despite the apparent 
absence of new stochastic contributions in $\Delta a_\text{MD}(t)$.
Furthermore, the two summands in Eq.~\eqref{eq:formal_solution} contribute in distinct ways, depending on 
whether the system is thermostatted or evolves in the same or similar way as in the microcanonical ensemble. 
Therefore, the variances associated with these contributions 
must be analyzed separately. In what follows, we focus on the thermostatted case.

Assuming a proper thermostat is present~\cite{Agarwal25JCP}, the function $C(t)$ will decay over time, 
and asymptotically, only the term involving the Green's function on the right--hand side of Eq.~\eqref{eq:formal_solution} will remain.
We refer to this remaining noise term as $g(t)$. Once $C(t)$ in Eq.~\eqref{eq:formal_solution} 
has effectively decayed to zero, $t \gg \tau$, a contribution to the virial estimator can be expressed as,
\begin{eqnarray}
\label{eq:F_a_times_a}
F_s a_s & = & K \left[ a_s^\text{eq}(h)  + g(t) - a_s^0 \right] \left[a_s^\text{eq}(h) + g(t)\right].
\end{eqnarray}
To facilitate the treatment, we restrict the remaining discussion to a stress--free reference, 
in which case $a_s^\text{eq}(h^\text{ref}) = a_s^0$ 
and $a_s^\text{eq}(h^\text{str})=(1+\varepsilon)a_s^\text{eq}(h^\text{ref})+ \varepsilon a_s' = a_s^0+\varepsilon a_0'$, with $a_s'$ from Eq.~\eqref{eqn:4b}.
It follows that
\begin{equation}
{F_s a_s}
  = K \left[ \varepsilon a_0' + g(t) \right]
\left[ a_s^0 + \varepsilon a_0' + g(t)\right] + \mathcal{O}(\varepsilon^2)
\end{equation}
so that the difference expression with an unstrained reference becomes
\begin{equation}
\label{eq:virial_product_difference}
\Delta (F_s a_s) = 2 a_0' \varepsilon [a_s^0 + g(t)] + \mathcal{O}(\varepsilon^2),
\end{equation}
where $a'_0$ denotes the change of equilibrium bond length with $\varepsilon$ relative to the (un--stressed) reference.
Eq.~\eqref{eq:virial_product_difference} shows that the leading–order ratio of noise $g(t)$ to signal $a_s^0$ is independent of $\varepsilon$.

When the difference between the virial in the strained sample and the (exact) expectation value of the virial is taken in Eq.~\eqref{eq:virial_product_difference}, an additional term proportional $g^2(t)-\langle g^2\rangle$ arises, which does not vanish as $\varepsilon$ tends to zero.
This term causes poor signal--to--noise ratio when noise cancellation is not used. 

\begin{figure}[hbtp]
\includegraphics[width=0.47\textwidth]{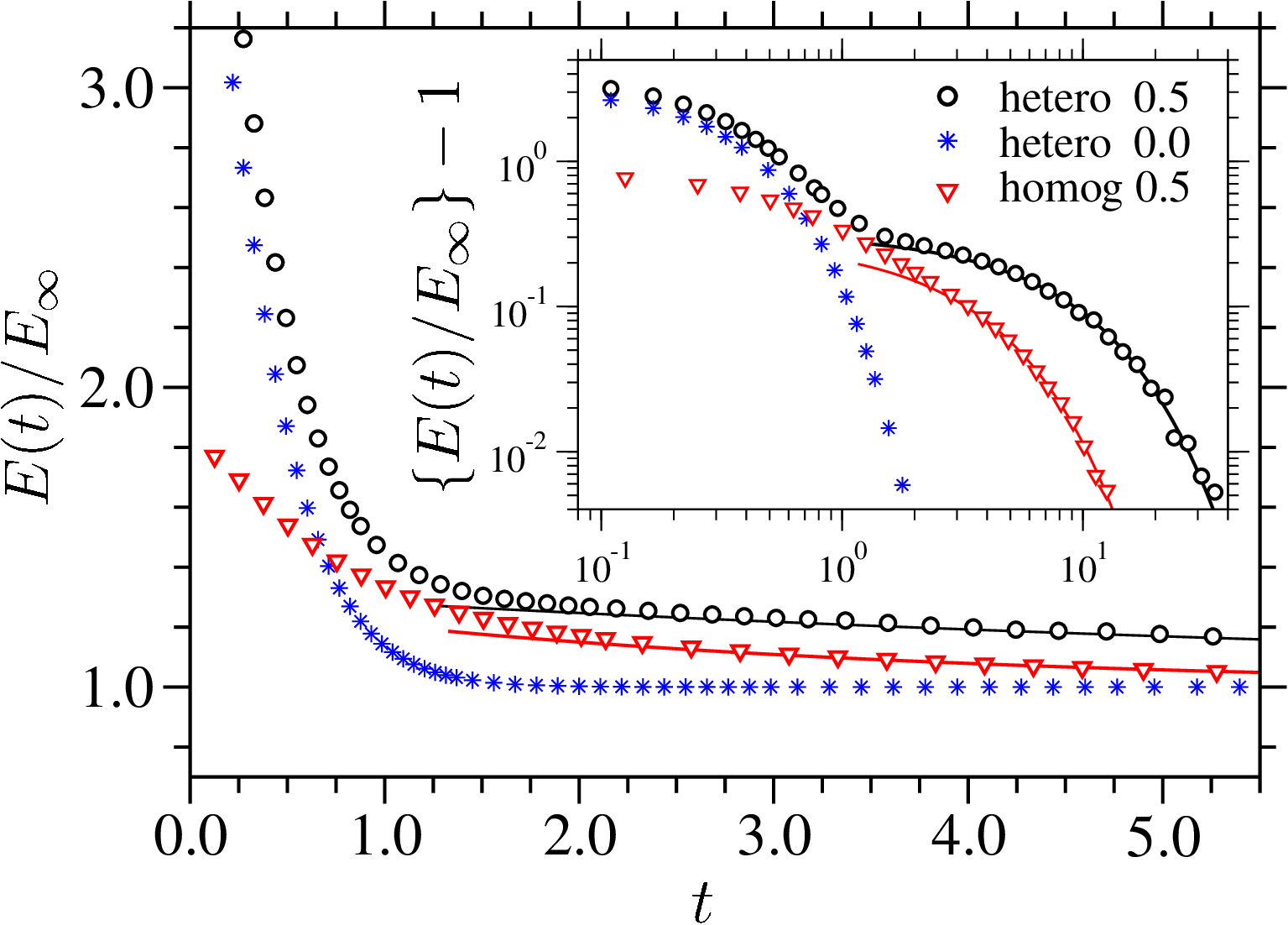}
\caption{Relaxation of the modulus $E(t)$ in a linear, ideally harmonic chain composed of three diatomic molecules. 
Three cases are shown: a heterogeneous (hetero) chain with alternating stiff, short bonds and soft, long bonds at finite 
temperature (black circles) and at zero temperature (blue stars), and a homogeneous (homog) chain with only one bond type. 
In all cases, the static modulus is $E_\infty = 1$. The inset displays the relative error of the estimate.
Solid lines indicate fits using a single-exponential relaxation model, $E(t) = E_\infty + \Delta E \exp(-t/\tau)$.}
\label{fig:HO_relax_all}
\end{figure}

To substantiate our calculations made in this section, Figure~\ref{fig:HO_relax_all} compares the time--dependent 
estimate for the modulus of linear, ideally harmonic chains. To this end, a heterogeneous and a homogeneous chain are considered.
The heterogeneous chain has short bonds ($a_s = 0.9$, $k_s = 10$) and long bonds ($a_l = 1.1$, $k_l = 0.52631\ldots$), 
where the exact numerical value for $k_l$ is chosen such that the static elastic modulus is  $E_{\infty} = 1$.
For the homogeneous chain, $a_s = a_l = 1$ and $k_s = k_l = 1$, also yielding $E_{\infty} = 1$.
The heterogeneous chain is simulated at finite temperature ($T = 0.5$) and at zero temperature, 
while the homogeneous chain is only investigated at $T = 0.5$.

For the zero--temperature heterogeneous chain, the elastic modulus estimate quickly reaches its asymptotic value because the non--affine deformation, namely the stiff optical $\Gamma$--point mode, relaxes rapidly.
In contrast, the finite--temperature {homogeneous} chain takes much longer to equilibrate, despite the absence of nonaffine (athermal) deformations.
This is because of low--frequency, long--wavelength modes.
In fact, a single (overdamped) mode, i.e., the one at the Brillouin--zone boundary, dominates the elastic--constant 
estimator at large times, as can be seen from the single exponential indicated by the corresponding full line in Figure~\ref{fig:HO_relax_all}.
The finite--temperature, heterogeneous chain displays both a fast initial relaxation due to the non--affine 
deformation and a slow final relaxation caused by the equilibration of long--wavelength acoustic modes.
Note that the latter would add to the relaxation function $C(t)$, but are absent in the 2--body system.

\subsection{Noise cancellation in quasi--harmonic systems}
\label{sss:qharm}

Interactions in real systems are never purely harmonic.
Even simple model potentials, such as the Lennard--Jones potential, contain anharmonic contributions.
Assuming a solid is nevertheless close to being harmonic, the potential energy can be approximated as
\begin{equation}
V(\varepsilon, \{u \}) \approx V_0(\varepsilon) + \sum_n \frac{k_{e,n}(\varepsilon)}{2} u_{e,n}^2,
\label{eq:quasi-harmonic}
\end{equation}
where the anhormonicity is represented by the dependence of the Taylor coefficients on strain. 
The index $e$ means to imply \textit{eigen} so that $k_{e,n}$ is the stiffness associated with the $n$--th eigenmode $u_{e,n}$.
In this quasi--harmonic approximation, the energy is truncated after the harmonic term, though it might be 
pragmatic to use effective values for $V_0(\varepsilon)$, which is the coefficient of interest, and $k_{e,n}$, e.g., 
to reproduce correctly the speed of sound in the continuum limit, rather than the ``lateral" expansion coefficients. 

In the realm of the noise--cancellation technique, we are primarily interested in deducing a difference of virial, which is dominated by the term $\Delta V = V_0(\varepsilon) - V_0(0)$ in a quasi-harmonic system. 
The difficulty that now arises in the quasi--harmonic approximation is that the harmonic contribution to the energy (difference) 
and thus the quasi--harmonic contribution to the stress and elastic--tensor estimators decorrelate over time, since the eigen\-stiffnesses, 
and thereby the eigenfrequencies $\omega_{e,n}$, depend on the strain.
In lowest order, $\omega_{e,n}(\epsilon) = \omega_{e,n}(0) + \omega'_{e,n} \varepsilon$.
Ignoring the effects of (implicit or explicit) damping, the difference between the \textit{harmonic} potential energy terms caused by 
a (random) momentum transfer $\Delta p$ at time $\Delta t$ after impact will be
\begin{eqnarray}
\Delta V_2 & = & \!
\frac{\Delta p^2}{2 m_e }\left[
\sin^2\{\omega_e(\varepsilon)\Delta t \} -
\sin^2\{\omega_e(0)\Delta t \}
\right]
 \label{eq:decoherence} \\
& = & 
\frac{\Delta p^2}{2 m_e }
\varepsilon \omega_e' \Delta t \sin \{2\omega_e(0)\Delta t\} + \mathcal{O}(\varepsilon^2).
\end{eqnarray}
Eq.~\eqref{eq:decoherence} indicates that the stress estimators, which are proportional to $\Delta V_2/\varepsilon$, 
begin to deviate at a rate independent of $\varepsilon$.
Initially, the deviation between the potential energy in the strained and reference samples grows as $\Delta t^2$, 
as long as $\Delta t$ remains small compared to the relevant phonon periods.
Once $\Delta t$ exceeds $1/\omega_e(0)$ but is still smaller than $1/\varepsilon \omega_e'$, the deviation grows linearly in time. 
However, a linear regime should not be expected in extended systems due to the broad distribution of eigenfrequencies.
Because the (instantaneous) stress, and hence the elastic--constant estimator, scale with $\Delta V_2/\varepsilon$, 
the time dependence of the spurious contribution is, to leading order, independent of $\varepsilon$.
This contribution is disadvantageous because 
it masks the genuine relaxation signal of self--affine modes.

In the limit of very large times, i.e., when $\varepsilon \omega_e'\Delta t \gg 1$, the two sinusoidal terms on the 
r.h.s. of Eq.~\eqref{eq:decoherence} can be viewed as two independent random numbers 
with identical mean and each having a standard deviation (std) of $1/2$.
Thus, the std of the \textit{instantaneous} stress estimator will diverge as $1/\varepsilon$.
This, however, does not imply that time averages suffer from the same problem.
In fact, the ``strained signal'' is a delayed ($\omega'_e<0$) or advanced ($\omega'_e > 0$) version of the 
reference signal, which leads to a systematic cancellation of integrated oscillations or fluctuations.
The smaller $\varepsilon$, the more closely in time the strained and reference signals coincide.
As a consequence, the std of an integrated signal decreases inversely with the observation time, $1/t_\textrm{obs}$.
If the strained signal were obtained from a completely independent run, this scaling would be limited by what might 
be conveniently called the curse of stochastic sampling, i.e., reducing the error by a factor of two requires quadrupling the number of samples.

So far, we have discussed how strained and unstrained modes decorrelate in the absence of damping.
This discussion thus pertains to the case when the damping is very small.
In the opposite limit of large damping, the response to an instantaneous (thermal) force is suppressed before the modes decorrelate.
This will lead to a finite, instantaneous std of the difference in the harmonic energy in the strained and the reference sample. 
This std will then remain independent of $\varepsilon$, even at large $t$.
Thus, large damping is preferred to tame stochastic errors at large times, which, however, causes the deterministic relaxation to be slow. 

Arbitrary damping can be dealt with in a straightforward fashion by using the exact Green's function and thermal random forces, 
i.e., second moments of the exact $\Delta V_2$ by solving the Langevin equation, squaring the results, and finally summing over eigenmodes and taking expectation values.
Instead of executing the just mentioned procedure, 
we find it more instructive to rationalize the behavior of a linear chain made, in which adjacent atoms interact through a Lennard--Jones potential.
Such an analysis also contains non--linear terms and coupling between modes. 

To this end, we consider a one--dimensional diatomic Lennard--Jones (LJ) chain consisting of three dimers with nearest--neighbor interactions. 
As in previous cases in Section~\ref{sss:noinse_harm}, the LJ length scales for the short and long bonds are 
chosen as $r_{0,s} =0.9$ and $r_{0,l} = 1.1$, respectively.
The LJ energy parameter $U_0$ is selected to satisfy the stiffness criterion: $k_{\rm 1d} = 72 U_0/r_0^2$.
This expression corresponds to the curvature of the 6--12 LJ potential at its minimum, located at $2^{1/6}r_0$. 
The value of $k_{\rm 1d}$ is chosen such that the zero--temperature, zero--stress reference state of the LJ chain 
matches the harmonic reference system previously analyzed in Section~\ref{sss:noinse_harm}, i.e., the one defined by 
$k_s = 10$ and $k_l = 0.52631...$

To introduce large thermal fluctuations, we increase the temperature to $T = 0.5$.
In this case, the atoms then go deeply into their repulsive interaction, leading to a modulus much in excess of 
the harmonic modulus where we increase the lattice constant of the reference system to $a_s + a_l = 3/2$, at which point $E$ is still greater than unity.
This choice might be seen as \textit{ad--hoc}, in particular as one--dimensional systems with short--range 
interactions can be seen as artefactual, e.g., they lack phase transformations.
However, sampling is extremely efficient and using the above choices for the one--dimensional LJ chain reproduces 
the trends of the full three--dimensional systems quite well. 
In this sense, the analysis of the one--dimensional LJ chain is meant to be a stepping stone between quasi--harmonic theory for a single 
Brillouin zone boundary vibration and a full three--dimensional system. 

\begin{figure}[ptb]
\includegraphics[width=0.47\textwidth]{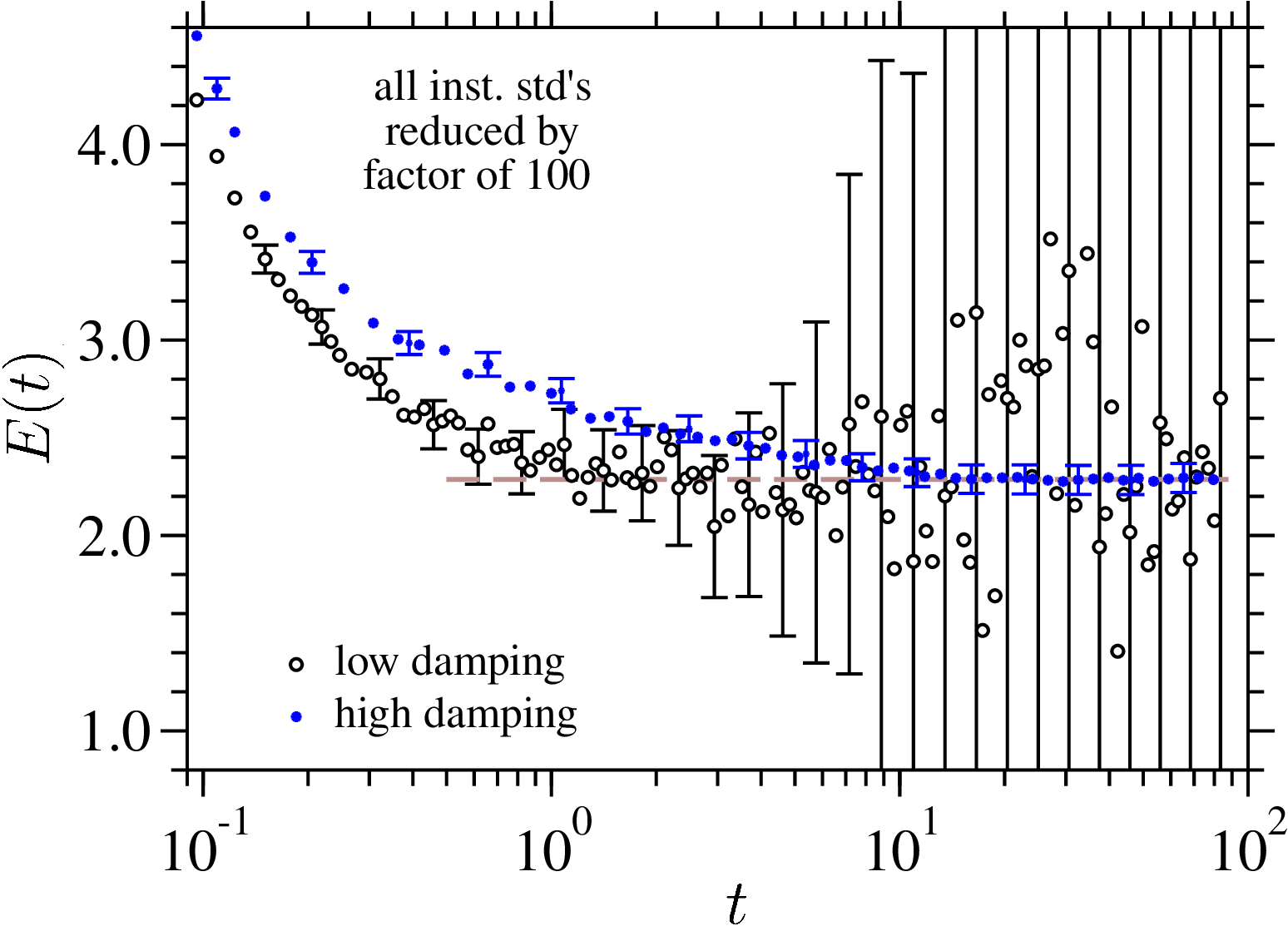}
\caption{\label{fig:LJ_relax}
Relaxation of the elastic--tensor estimator for a linear Lennard--Jones chain using the noise--cancellation technique with $\varepsilon = 10^{-3}$, 
shown for two relaxation times: $\tau = 0.109$ (large damping) and $\tau = 0.436$ (small damping).
Averages were taken over 200,000 samples. Instantaneous error bars (std) have been reduced by a factor of 100 for clarity.}
\end{figure}

Figure~\ref{fig:LJ_relax} confirms similar trends for the one--dimensional LJ chain as those just discussed in Section~\ref{sss:noinse_harm}.
In particular, weaker damping leads to a faster relaxation of the expectation value for the elastic--constant estimator 
than large damping but also to more quickly growing variances. 

The results for the linear LJ chain suggest that choosing thermostat time constants is not entirely straightforward.
While it is tempting to formulate general guidelines, our attempts proved of limited value for the complex systems analyzed in the results section.
In practice, trial and error was the most effective way to balance the slow initial relaxation caused by large damping against 
the large stochastic errors that arise at large simulation times with small damping.

\begin{figure}[ptb]
\includegraphics[width=0.47\textwidth]{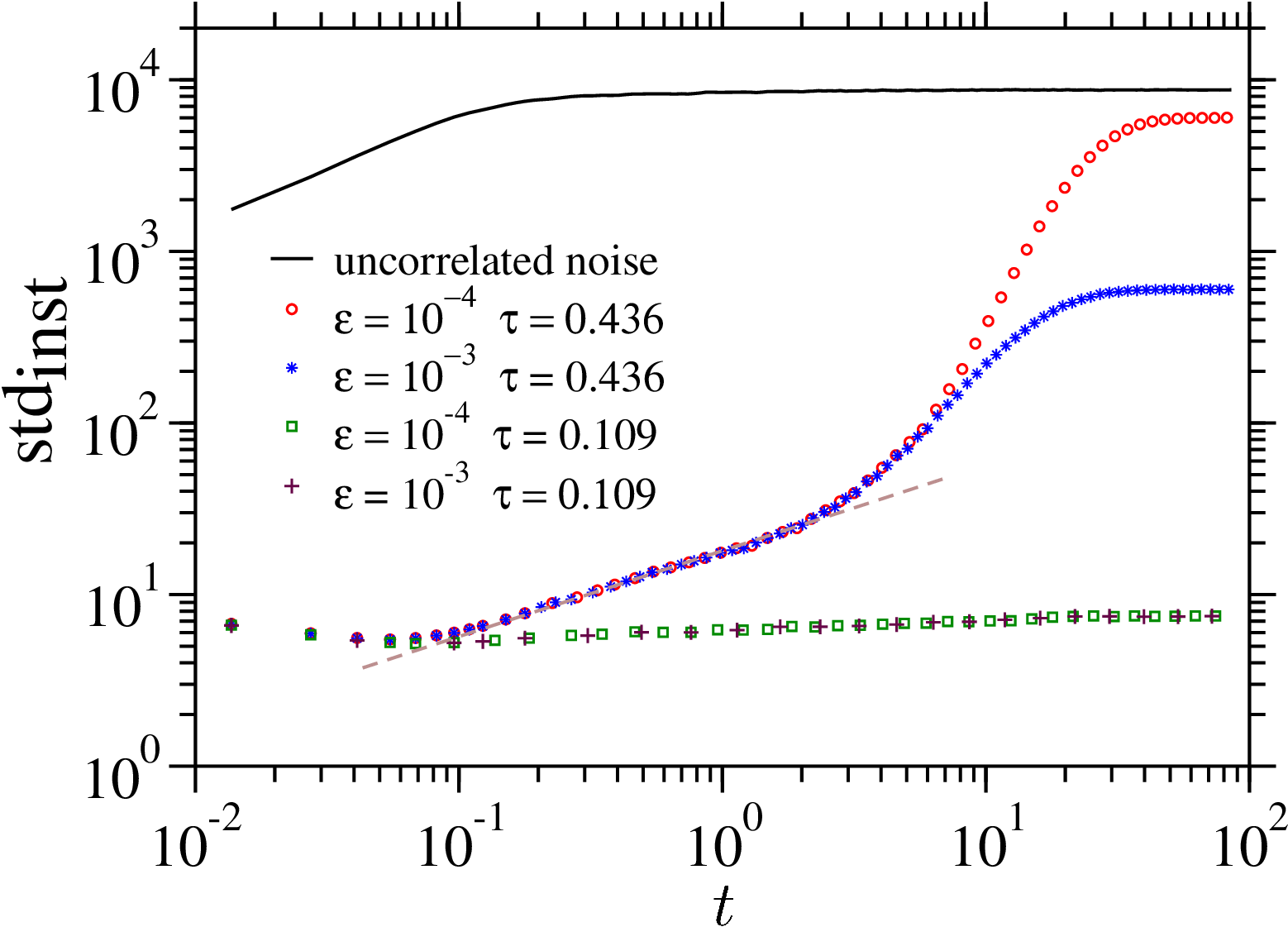}
\caption{\label{fig:LJ_std}
Instantaneous standard deviation for data like that shown in Figure~\ref{fig:LJ_relax}. 
In the calculation using uncorrelated random numbers for the follow--up runs, a strain of $\varepsilon = 10^{-4}$ was 
imposed and large damping $\tau = 0.109$ was used. The dashed line shows a $\sqrt{t}$ power law. 
}
\end{figure}

While Figure~\ref{fig:LJ_relax} already confirmed the expectation that larger damping implies smaller standard deviation 
of the instantaneous measurements, Figure~\ref{fig:LJ_std} corroborates this result using a double--logarithmic 
representation of the standard deviation as a function of time. 
This analysis also shows that the moment in time, when the standard deviation starts to increase does not depend on the strain $\varepsilon$, 
at least as long as it is sufficiently small.
%
%
Moreover, after transiting through another regime, the std levels off at a larger final value as the strain decreases.
Despite being large, the std remains far below the stochastic errors that one would obtain if the two follow--up simulations were run using a different set of random numbers.
Perhaps, most importantly, when using sufficiently large damping, the long--time standard deviation is independent in leading order on $\varepsilon$.

Higher-order terms in the potential energy are not expected to change the results qualitatively if treatable by perturbation theory.
Any resulting corrections to the estimators scale with $\varepsilon$ or higher, leaving the leading-order behavior intact.

Fig.~\ref{fig:LJ_std} reveals that standard deviations can be large with non-optimized parameter choices. 
We thus explored the performance compared to the `traditional' method using the Born term and a stress-fluctuation correction~\cite{Lutsko1989JAP} for the linear LJ chain. 
We found a wide spread of results, with standard deviations ranging from being on par to being reduced by a factor of 100 when using the noise cancellation technique. 
However, while Born terms are readily evaluated for pair potentials, this is not true for many other potentials. 
Thus, the important message of Fig.~\ref{fig:LJ_std} is that the noise cancellation technique yields smaller stochastic errors than uncorrelated differences.

\section{Materials, models, and numerical methods}
\label{sec:method}

\begin{figure}[ptb]
\includegraphics[width=0.49\textwidth]{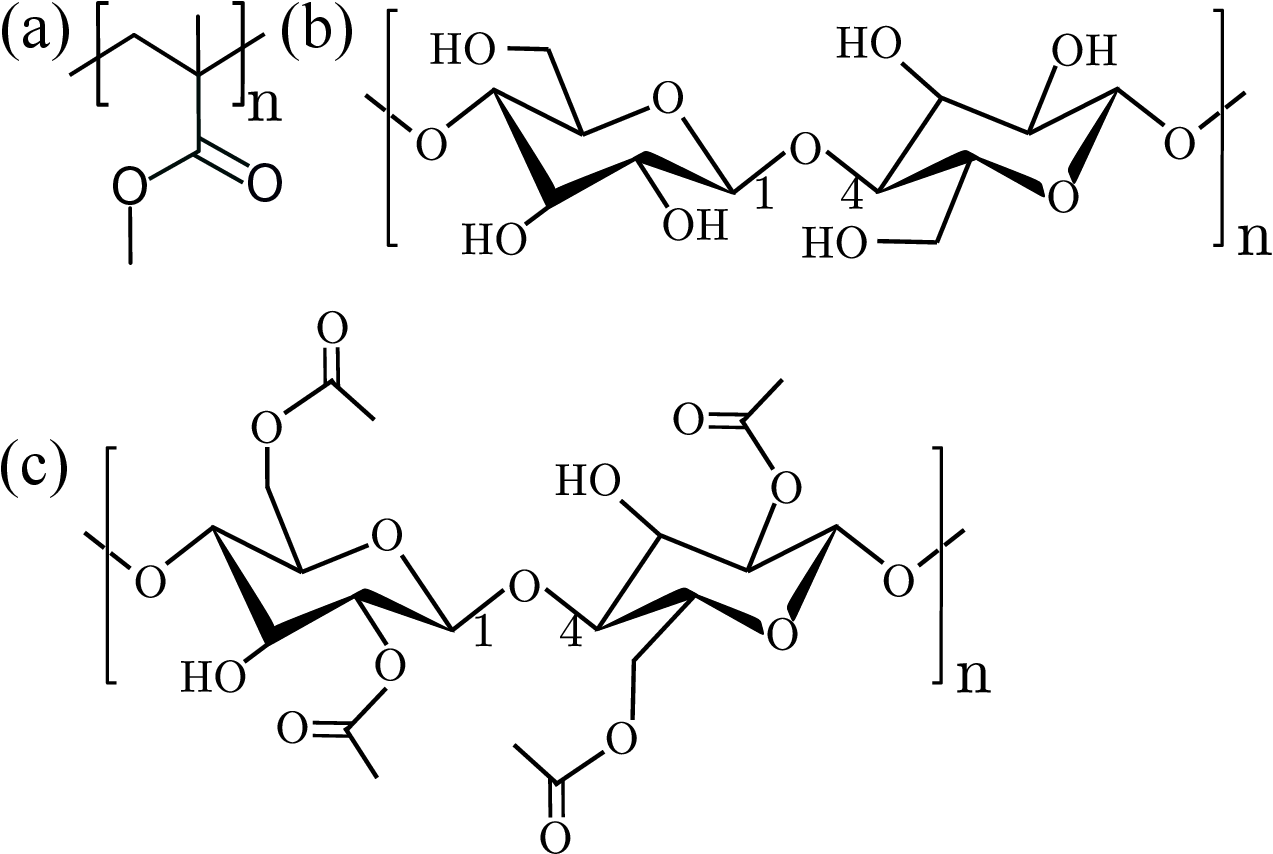}
\caption{Monomer structures of poly(methyl methacrylate) (part a), cellulose (part b), and cellulose acetate (part c)
ivestigated in this study.}
\label{fig:schem}
\end{figure}

We have tested the proposed method across a range of realistic models with increasing 
degrees of structural complexities, beginning with face--centered cubic argon, 
followed by crystalline silicon in the diamond structure and amorphous silicon, and 
concluding with amorphous polymers, specifically poly(methyl methacrylate) (PMMA), cellulose, and 
cellulose acetate. Schematic representations of the monomer structures of these polymers are shown in Figure~\ref{fig:schem}.
The simulations for argon and silicon were performed using the LAMMPS package~\cite{thompson2022lammps}, 
while polymer simulations were carried out using GROMACS~\cite{Abraham:2015}.

\subsection{Argon}

In the argon simulations, individual particles interact via the standard 6--12 Lennard–Jones (LJ) potential, 
with an interaction energy of 0.2381 kcal/mol (10.32 meV) and a characteristic LJ length scale of 3.405~{\AA}~\cite{Rahman64Ar}. 
A cut--off distance of 9.0~{\AA} is applied to truncate the interactions.
The initial atomic configuration is generated as a face--centered cubic (FCC) lattice consisting of 
500 atoms, using a lattice constant of $a = 5.311$~{\AA}~at $T = 10$ K~\cite{Barrett64JCP}.
The elastic tensor components $C_{ij}$ are computed at a temperature of $T = 10$ K over a total simulation time of 
10 ps, using a time step of $\Delta t = 1$ fs.
Temperature control is achieved using a Langevin thermostat with Gr{\o}nbech--Jensen algorithm~\cite{GronbechJensen13}, 
with a relaxation time constant of $\tau_{\rm T} = 0.1$ ps. We note that the Debye temperature of 
crystalline argon is approximately 82 K; thus, the simulations are carried out well within the 
regime where the harmonic approximation remains valid in a classical treatment.
For comparison, additional simulations were performed using the 
canonical sampling through velocity rescaling (CSVR) thermostat~\cite{Parrinello07jcp} and
the microcanical ensemble.

\subsection{Silicon}

Silicon simulations are conducted in two stages: (i) crystalline and (ii) amorphous configurations. 
In both cases, the systems consist of 1,000 silicon atoms. Atomic interactions are modeled using the 
modified Tersoff potential~\cite{OriginalTersoff1988,ModTersoff07}.

The crystalline sample is initialized in the diamond cubic structure, using a lattice constant of $a = 5.431$~{\AA}~\cite{Griesser23prm}. 
Starting from this configuration, the amorphous melt is generated by heating the system to 
$T = 4{,}000$ K to induce melting. The resulting melt is then quenched to 300 K at a cooling rate of 
1.48 K/ps using $\Delta t = 5$ fs and $\gamma_{\rm T} = 1.0$ ps.
During this stage, a constant pressure of $p = 1$ bar is maintained 
using the Parrinello--Rahman barostat.
The elastic tensor components $C_{ij}$ are computed at $T = 300$ K, with temperature controlled via a 
Langevin thermostat with Gr{\o}nbech--Jensen algorithm~\cite{GronbechJensen13}, using a time constant $\tau_{\rm T} = 0.01$ ps. 
These simulations are typically run for 10 ps with a time step of $\Delta t = 1$ fs. 
In some cases, the convergence of $C_{ij}$ is also tested with different $\tau_{\rm T}$; such details are reported where relevant.
We have also tested alternative quenching protocols for preparing the amorphous silicon glass. 
The corresponding details are provided in Supplementary Section S1~\cite{epaps}.

\subsection{Poly(methyl methacrylate)}
\label{ss:pmma}

As a representative example of an amorphous commodity polymer, we simulated a sample of poly(methyl methacrylate) (PMMA). 
The PMMA is modeled using modified OPLS--AA parameters~\cite{Mukherji17NC}, which have previously been shown 
to accurately capture the conformational behavior of single PMMA chains in solutions~\cite{Mukherji17NC}, 
as well as the heat capacity~\cite{MHM21prm} and thermal conductivity~\cite{Mukherji24PRM} of solvent--free bulk PMMA.
For this study, we employed a pre--equilibrated PMMA melt sample generated in earlier work~\cite{MHM21prm}. 
The system comprises 100 PMMA chains, each with a degree of polymerization of $N_{\ell} = 30$.
This leads to 45200 atoms in a PMMA sample.

The sample was equilibrated in the melt state at $T = 600$ K for at least $2~\mu$s, followed by quenching to $T = 300$ K~\cite{Mukherji24PRM}, 
at which point the elastic tensor components $C_{ij}$ were computed.
Temperature control was achieved using a Langevin thermostat, and the simulations were carried out 
for 200 ps with a time step of $\Delta t = 0.4$~ps. 
The thermostat coupling time was set to $\tau_{\rm T} = 0.1$ ps for the production runs. 
Additionally, various damping schemes were implemented for the PMMA sample studied in this work; specific details are provided where relevant.

Long--range electrostatic interactions are handled using the particle--mesh Ewald (PME) method~\cite{pme}, 
and non--bonded interactions are truncated at a cutoff distance of 
$r_{\rm c} = 1.0$ nm.

\subsection{Cellulose and cellulose acetate}

We used pre--equilibrated amorphous samples of cellulose and cellulose acetate, previously 
generated in our recent study~\cite{MukherjiMac26}. Both systems consist of 601 chains, each with 
a degree of polymerization $N_{\ell} = 50$. 
Cellulose and cellulose acetate samples consist of about 0.63 and 0.93 million atoms, respectively.
To obtain well--equilibrated amorphous structures, the cellulose melt was equilibrated for $4.5~\mu$s and 
cellulose acetate for $2.5~\mu$s at $T = 800$ K. After equilibration, both systems were gradually cooled to 
$T = 300$ K, at which point the elastic tensor components $C_{ij}$ were computed.

Cellulose was modeled using the GLYCAM06 force field~\cite{GLYCAM08,Glycam19}, which reliably 
captures the chair conformation of D--glucose rings forming the cellulose backbone. 
Simulations of the cellulose systems followed the same protocol as used for PMMA.

\subsection{Computation of elastic tensor elements}

In this work, we focus on the tetragonal shear modulus
\begin{equation}
G = \frac{C_{11}-C_{22}}{2}.
\end{equation}
The reason for this choice is that shear moduli are smaller than bulk moduli and thus generally 
suffer from poorer signal to noise ratios, which puts large demands on our method. 
Moreover, the original motivation for this work stems from one of the authors' (DM) interest 
in computing high--precision shear modulus data for amorphous polymers using GROMACS~\cite{MukherjiMac26}. 
Given that the cellulose samples contain approximately 0.63--0.93 million all--atom particles -- 
and considering the known limitations of fluctuation--based methods -- 
we compute the tetragonal shear modulus indirectly by applying a volume--conserving strain to the system via;
\begin{subequations}
\begin{eqnarray}
    \varepsilon_1 & = & \varepsilon \\
    \varepsilon_2 & = & \frac{1}{1+\varepsilon}-1 \\
    \varepsilon_{i\ge 3} & = & 0.
\end{eqnarray}
\end{subequations}
In isotropic materials, this tetragonal shear modulus is equivalent to the conventional shear modulus.
Approximating $\varepsilon_2 \approx 1 - \varepsilon$ for the small $\varepsilon$ values, we thus find for systems of cubic or higher symmetry
\begin{equation}
\sigma_{1,2} = \pm (C_{11}-C_{12}) \varepsilon
\end{equation}
so that $G$ can be deduced in leading order via
\begin{equation}
\label{eq:shear_modulus_estimator_1}
    G = \frac{\sigma_1-\sigma_2}{2\varepsilon}.
\end{equation}

Other elements of the elastic tensor could be deduced through related box--deformations, e.g., $C_{11}$ by leaving all 
strain--tensor elements zero, except for $\varepsilon_1$.
In this case, noise cancellation might best be done by deducing $C_{11}$ from taking the difference signal of two strained simulations via
\begin{eqnarray}
C_{11} & = & \frac{  \sigma_1(\varepsilon_i=\varepsilon\delta_{i,1}) - \sigma_1(\varepsilon_i=-\varepsilon\delta_{i,1}) }{2\varepsilon} \nonumber \\
C_{12} & = & \frac{  \sigma_2(\varepsilon_i=\varepsilon\delta_{i,1}) - \sigma_2(\varepsilon_i=-\varepsilon\delta_{i,1}) }{2\varepsilon}
\label{eq:c11_c12}
\end{eqnarray}
Ultimately, by choosing various ``shapes'' for the strain tensor, all $C_{ij}$ can be determined.
By considering different magnitudes in the strains, the systematic expansion of the (free) energy beyond quadratic terms can be targeted. 

\section{Results}
\label{sec:res}

To further detail the method and to substantiates claims made in Section~\ref{sec:ThMe}, 
the time evolution of elastic modulus estimators in Eq.~\eqref{eq:shear_modulus_estimator_1} and Eq.~\eqref{eq:c11_c12} will be scrutinized 
for selected samples using selected approaches.

\begin{figure}[btp]
\includegraphics[width=0.49\textwidth]{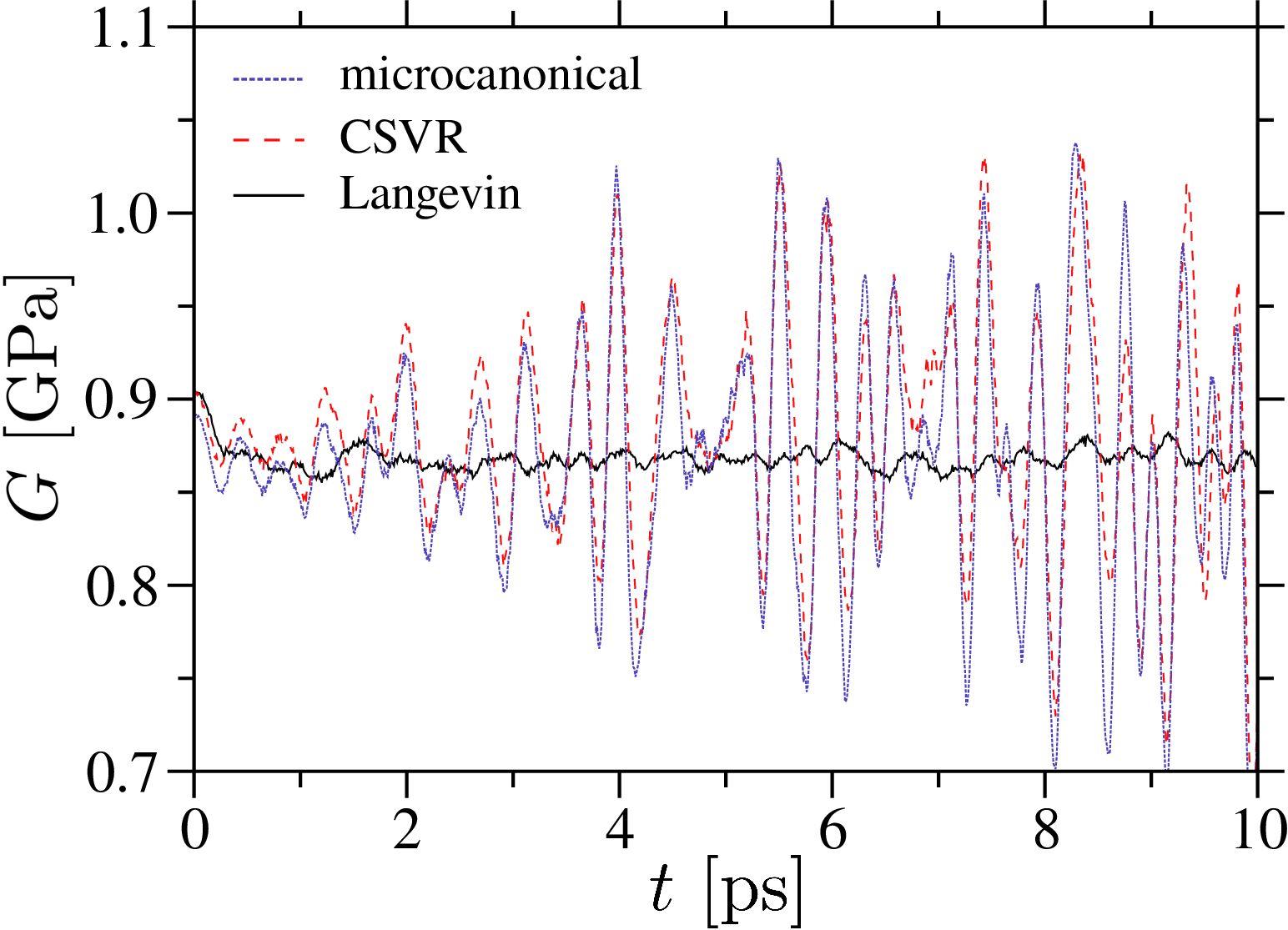}
\caption{Time evolution of the tetragonal shear--modulus estimator defined in Equation~\eqref{eq:shear_modulus_estimator_1} 
for crystalline argon at a temperature $T = 10$ K. Different integration schemes are used: microcanonical (blue, dotted line), 
canonical--sampling by velocity rescaling (CSVR, red, dashed lines), and 
Langevin (full, black line) using the Gr{\o}nbech--Jensen algorithm.}
\label{fig:cryst_ar_therm}
\end{figure}

\begin{figure}[btp]
\includegraphics[width=0.49\textwidth]{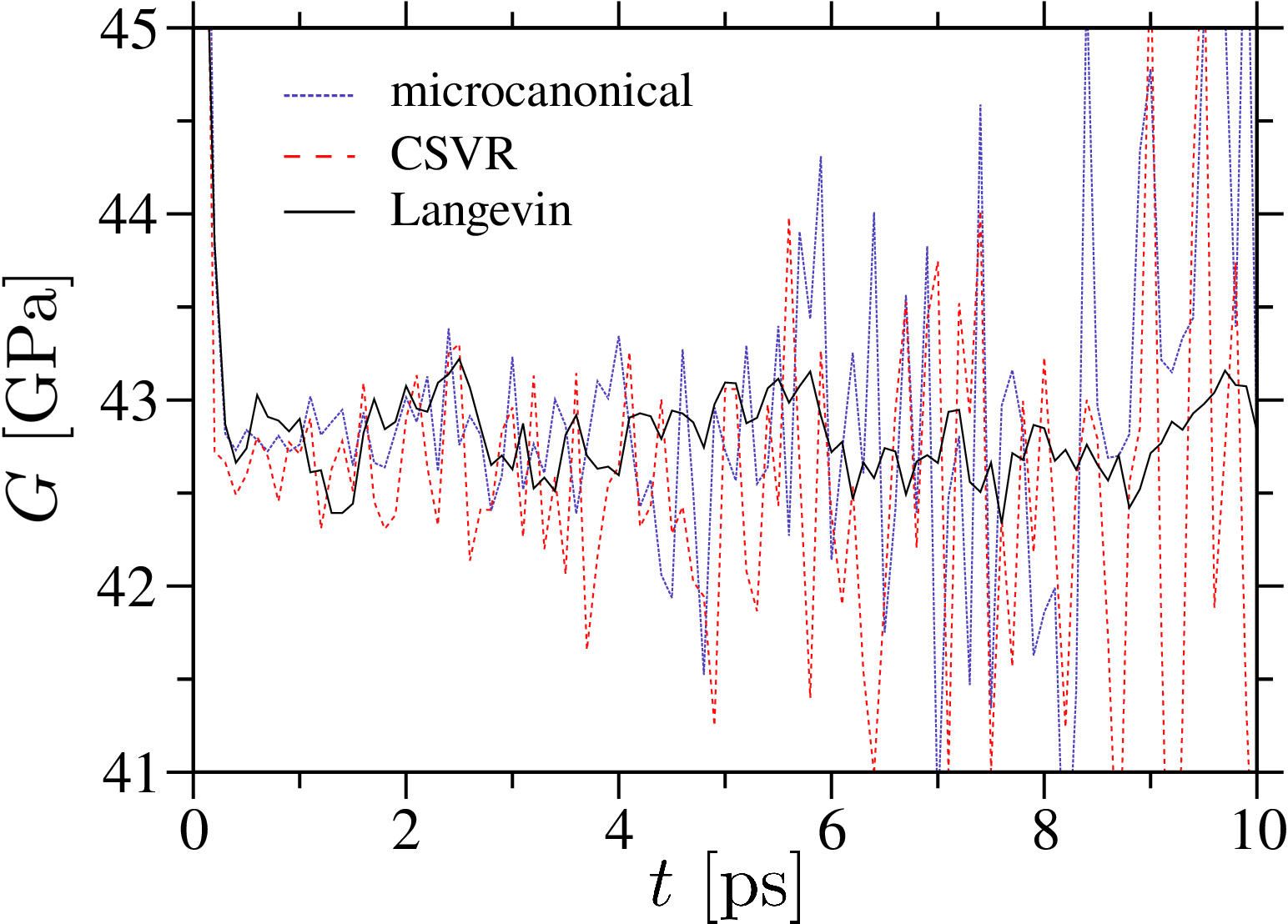}
\caption{Same as Figure~\ref{fig:cryst_ar_therm}, however, for silicon at a temperature $T = 300$ K.}
\label{fig:cryst_si_therm}
\end{figure}

\subsection{Effect of thermostat}

The calculation in Section~\ref{sss:noinse_harm} was based on the assumption that $C(t)$ in Eq.~\eqref{eq:formal_solution} has decayed.
Thermostatting should therefore aim to achieve this decay as efficiently as possible.
Therefore, we begin by examining the impact of thermostat choice on simulations of crystalline argon and silicon, 
as shown in Figure~\ref{fig:cryst_ar_therm} and Figure~\ref{fig:cryst_si_therm}.
Interestingly, the widely used canonical sampling through velocity rescaling (CSVR) thermostat \cite{Parrinello07jcp}-- 
a global kinetic energy control method-- produces dynamics that closely resemble those of the microcanonical (NVE) ensemble.
While all schemes appear to level off at a similar estimate for the shear modulus at very early times,
a notable difference, however, emerges in the later relaxation 
between thermostatted dynamics (solid black line) and pseudo--microcanonical schemes (dotted blue and dashed red lines). 
This difference arises because the microcanonical ensemble does not enable energy exchange between vibrational modes, which can limit convergence of the shear modulus, $G$,
over time-- particularly in nearly harmonic systems.
Increasing the system size reduces the stochastic noise. Additional simulations on crystalline silicon, 
in which the system size was increased from $N = 1,000$ to 8,000, suggest that this reduction obeys the usual curse of stochastic sampling, 
i.e., that system sizes have to be quadrupled to double the signal-to-noise ratio (data not shown).

Despite above discrepancies-- between the properly thermostatted and microcanoncal simulations-- 
in transient behavior, the time--averaged values of $G$ differ by less than 1.5\% 
across the different approaches. However, the associated statistical uncertainty-- estimated via block--averaged standard deviations-- 
exceeds 10\% for both the NVE and CSVR ensembles, while it drops to approximately 0.5\% when using the Langevin thermostat \cite{GronbechJensen13}.
Although similar trends are observed for both crystalline argon at $T =10$ K and silicon at $T = 300$ K (both well witin their respective harmonic limits,
i.e., significantly below the Debye temperature), 
the case of silicon is more nuanced due to non--affine displacements of the basis atoms in the diamond structure under anisotropic deformation.
Note that the quantum flcutuations only alter elastic constants well below the Debye temperature~\cite{Schoffel2001PRB,Martonak98PRE,Muser01jcp}.

We note in passing why the microcanonical setup exhibits large fluctuations. Ideally, in the absence of external noise, it 
should yield a well--converged value. However, as shown in Eq.~\eqref{eq:formal_solution}, the Green's function 
term, $g(t)$, vanishes, leaving only the first term, which barely decays over time without explicit damping. 

In addition to the system-size dependence of the signal-to-noise ratio, one may be concerned about finite-size effects in the signal itself.
For crystalline systems, this is typically minor, as sufficiently large simulation cells effectively provide a ``$k$-point sampling'' in the sense of the trapezoidal rule and the response is dominated by the $k=0$ contribution.
In macroscopically homogeneous but microscopically disordered systems, more pronounced finite-size effects must be expected.
In practice, corrections will become small once the linear simulation cell size exceeds at least twice the structural 
(Ornstein–Zernike) correlation length (typically around $1.5$~nm), and, in the case of polymers, the corresponding end--to--end distance.
With linear box sizes of 7.8~nm for PMMA~\cite{MHM21prm}, 18.0~nm for cellulose, and 21.6~nm for cellulose acetate~\cite{MukherjiMac26}, 
our simulations satisfy these criteria comfortably so that finite-size effects are expected to remain small.

\subsection{Effect of thermostat damping time}


\begin{figure}[btp]
\includegraphics[width=0.49\textwidth]{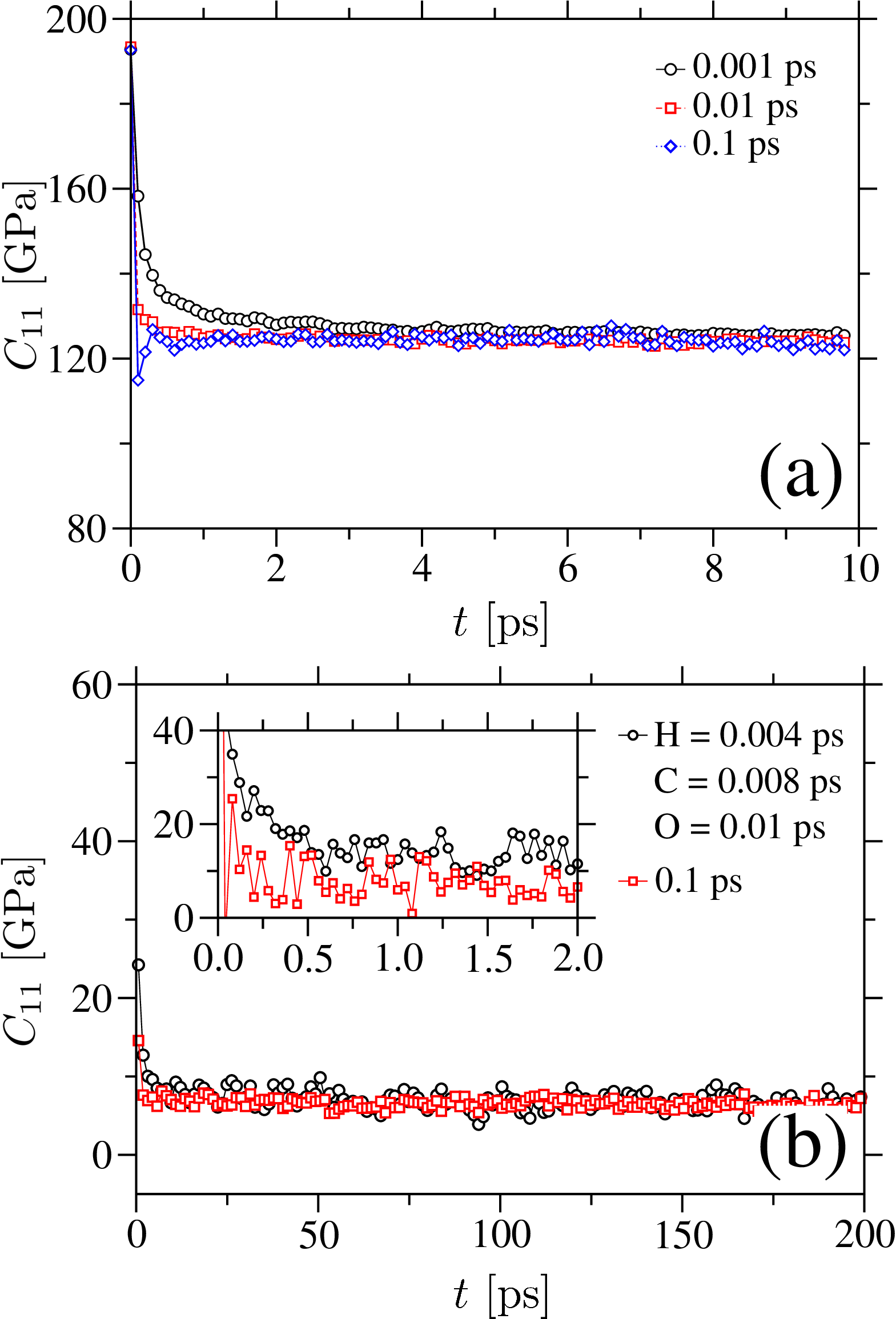}
\caption{The temporal relaxation of the elastic modulus $C_{11}$ is shown for two amorphous systems: 
amorphous silicon (panel a) and poly(methyl methacrylate) (PMMA) (panel b). The inset in panel (b) highlights the short--time 
relaxation behavior ($t \le 2$ ps) of $C_{11}$. Each panel displays data corresponding to different 
damping times of the Langevin thermostat. For clarity, the data in the main panel (b) have been coarse--grained, 
averaging every 32 data points into one.} 
\label{fig:C11_three}
\end{figure}

\begin{figure*}[ptb]
\includegraphics[width=0.997\textwidth]{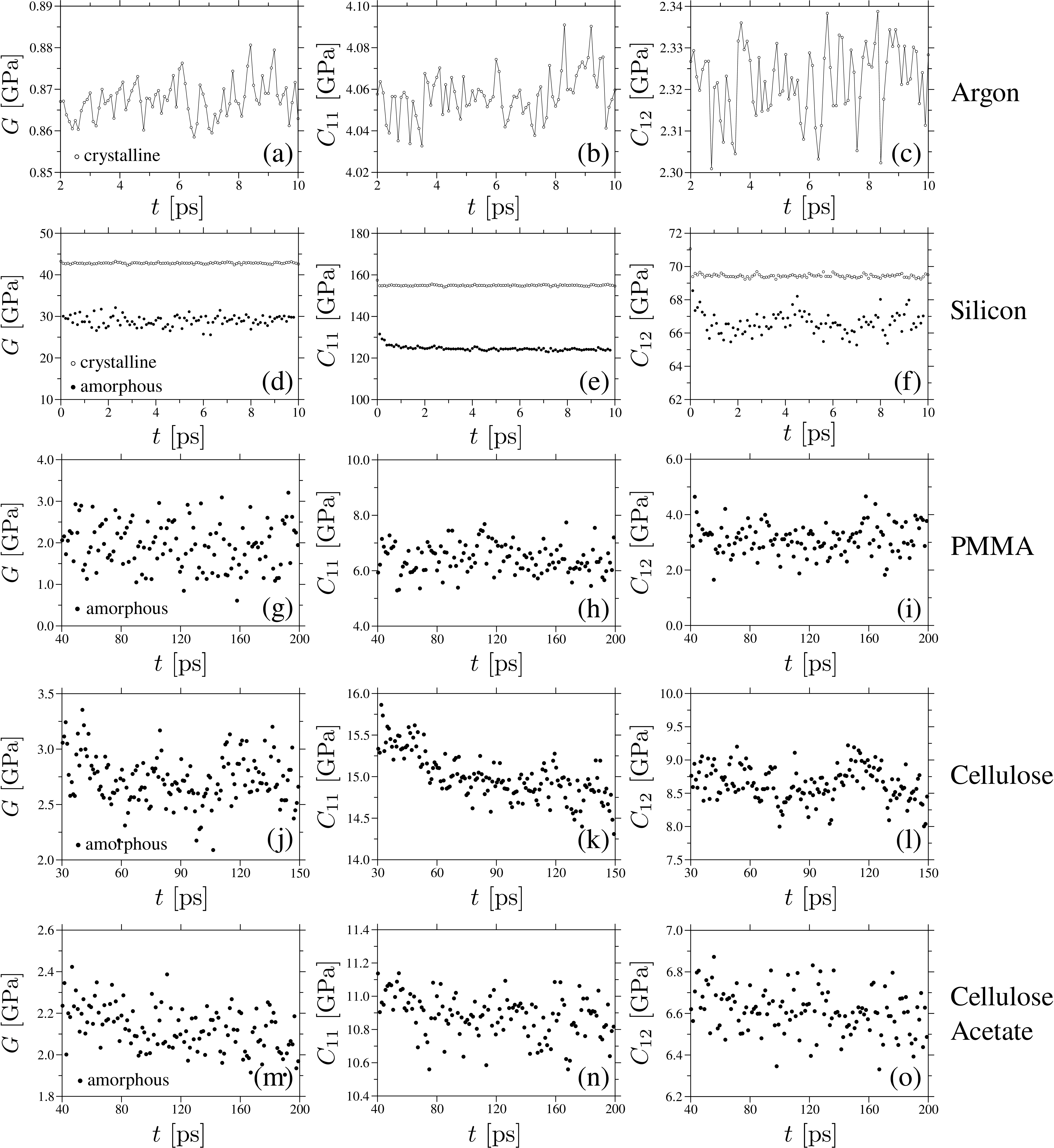}
\caption{Notably, the results for crystalline argon and silicon are averaged over 1 and 16 independent runs, respectively, while the amorphous silicon 
data represent an average over 49 runs. For all polymer systems, the reported values are averaged over 100 independent simulations.
The polymer data shown herein have been coarse--grained for clarity.}
\label{fig:cij_all}
\end{figure*}

In Figures~\ref{fig:LJ_relax} and~\ref{fig:LJ_std}, we demonstrated the influence of damping on the elastic 
modulus estimator for a one--dimensional Lennard--Jones chain. As discussed, the relaxation function $C(t)$ 
in Eq.~\eqref{eq:formal_solution} decays most rapidly when a mode is critically damped.
In real systems, however, the presence of a distribution of eigenfrequencies complicates this behavior. 
For practical equilibration, it is crucial to efficiently relax those modes that contribute most significantly to the decay of the initial excess stress.
Since non--affine deformations are typically localized-- at least in homogeneous materials-- 
the relevant vibrational modes tend to lie at higher frequencies. Therefore, it is advantageous to first target 
the high--frequency vibrational modes associated with stiff bonds, such as OH, CH, and CC bonds in polymers~\cite{MHM21prm}.

Since the stress in stiff covalent bonds is the dominant contribution to the excess stress after an affine deformation, 
it is meaningful to first use a relatively short thermostat time constant, such that the fastest modes become critically damped.
During this initial stage, most of the excess stress is rapidly reduced. However, other non--affine modes, 
such as those related to torsional motion, will be overdamped and thus hindered in their relaxation.
For this reason, it can be beneficial to follow the initial run, or, rather pairs of runs, with one or 
more subsequent runs using a substantially longer thermostat time constant.
The need for such runs and the optimal choice of parameters 
will generally depend on the specific system under study.
If this second phase is omitted, slow relaxation of the signal can persist, potentially leading to 
incorrect conclusions about the flow behavior of the system studied. 

Using accurate integration schemes, such as the one proposed by Gr{\o}nbech--Jensen~\cite{GronbechJensen13}, 
one can choose relatively large time steps such that the fastest quasi--harmonic mode can be relaxed within roughly few hundred MD steps-- 
though running a few more does no harm.
If the second stage targets modes with periods ten times longer than the fastest ones, approximately 
ten times more time steps will be needed, since the fastest modes still limit the timestep size.
In principle, the procedure can be iterated, and the scaling of the thermostat time constant can be optimized.
However, it may also be advantageous to conduct the follow--up simulation with the lowest--frequency mode critically damped. 
Ideally, the simulation should then run long enough for a transverse sound wave to traverse the simulation cell.

Further optimizations may be needed, especially since small damping produces large instantaneous 
(though not necessarily time--averaged) asymptotic standard deviations.
In principle, one would expect it to be beneficial if the broad spectrum of frequencies in 
non--elemental systems were somewhat collapsed if atoms participating particularly strongly in high--frequency vibrations were made heavier.
However, in the relevant cases investigated for polymers, the benefits appeared to be limited.
Instead it proved more useful to use shorter damping time constants for the light hydrogen atoms than for the heavier carbon atoms. 
The reason could be that torsional and other slow modes were less overdamped by this procedure.

Another optimization is to not apply an ideally affine strain.
For example, only the center of mass and/or the backbone of a polymer could be subjected to 
an affine transformation, while the remaining internal degrees of the polymers remain unaltered. 
In this case, ligands and side groups could be re--attached to the backbone atom using unaltered distance vectors.  

\subsection{Applications}
\label{sss:app}

Figure~\ref{fig:C11_three} presents results for two different samples, each simulated using a range of Langevin damping times 
$\tau_{\rm T}$. At small damping times-- where stiff vibrational modes are critically damped-- they equilibrate rapidly. 
However, this also leads to a slower overall relaxation of $C_{11}$. In contrast, larger damping times result in slower equilibration 
of high--frequency modes, but the average value around which they fluctuate converges more quickly to the long--time relaxed value of 
$C_{11}$. A trend that is also evident for the simplified model in Figure~\ref{fig:LJ_relax}.

We note in passing that the silicon system exhibits a relatively narrow range of characteristic vibrational frequencies,
where we have attempted one damping time.
However, for polymers-- such as PMMA-- which possess a broad spectrum of eigenfrequencies~\cite{MHM21prm}, 
we have also used various Langevin damping schemes with a goal to explore different possibilities to improve the data. 
These details are highlighted in the Supplementary Section~S1.

In Figure~\ref{fig:C11_three}(b), the PMMA data correspond specifically to damping schemes (i) and (ii) 
discussed in the Supplementary Section S1 for the elastic modulus $C_{11}$. A full set of PMMA results for all elastic constants obtained using these damping strategies is provided in the 
Supplementary Figure~S2~\cite{epaps}.

For the production runs of all systems studied in this work, we selected damping times, $\tau_{\rm T}$, that ensure rapid convergence of the elastic constants, $C_{ij}$, 
to their reference values. Added advantage of the single damping time scheme is that it is easy to implement even in the
complex polymeric systems. 
Figure~\ref{fig:cij_all} summarizes the time evolution and convergence behavior of $C_{ij}$ for each system.
Notably, after the initial rapid relaxation, the amorphous samples exhibit a secondary, slower decay in the elastic constants 
$C_{ij}$ over time. This effect becomes increasingly pronounced with greater structural complexity-- for instance, 
the decay is noticeably slower in amorphous cellulose compared to amorphous silicon (see Figures~\ref{fig:cij_all}e, k, and n).
A similar two--stage relaxation trend is observed in the case of the generic diatomic harmonic chain with thermal fluctuations 
(see Figure~\ref{fig:HO_relax_all}). At first glance, and in the absence of the theoretical insights provided by Figure~\ref{fig:HO_relax_all}, 
this transient behavior might be misinterpreted as an indication of flow or irreversible deformation within the sample.
To rule out this possibility, we conducted a series of additional simulations involving energy minimization on pre--strained 
cellulose samples. In these tests, the atomic coordinates remained unchanged after relaxation to the next energy minimum between the initial and final configurations (data not shown), thereby confirming the absence of any flow--like behavior.
Taken together, these observations suggest that the slow decay in $C_{ij}$ is not due to structural rearrangements 
but rather arises from intrinsic vibrational eigenfrequencies of the system. Specifically, the broad distribution of 
eigenfrequencies characteristic of polymeric and amorphous materials leads to a slow convergence of the elastic response. 
In particular, the overpopulation of low--frequency vibrational modes-- commonly referred to as the 
Boson peak~\cite{Leonforte2005PRB,HORBACH1998320}-- is a key factor underlying the long--time decay of $C_{ij}$.

\begin{table*}[ptb]
        \caption{Average values of the elastic modulus components.
They are computed by block--averaging over ten blocks taken from the second half of the data shown in Figure~\ref{fig:cij_all}. The standard errors are calculated across these ten blocks.
For comparison, available literature data for different systems are also included.
In all quoted experimental values, $G$ was obtained from  $G=\left(C_{11}-C_{12}\right)/2$.}
\begin{center}
       \begin{tabular}{|c|c|c|c|c|c|c|c|c|c|c|}
\multicolumn{5}{c}{Crystalline samples} \\
\hline
System & Source &  $G$ [GPa]  & $C_{11}$ [GPa]  & $C_{12}$ [GPa] \\
\hline
\hline
\hline
           Argon  & This work                        & $0.868 \pm 0.001$   &  $4.069 \pm 0.002$   &  $2.325 \pm 0.001$    \\
                  & Previous sim.~\cite{Schoffel2001PRB}  &      --             &   3.9                &    --                 \\
                  & Experiment~\cite{Luscher1972PL}  &      1.0            &   4.0                &  2.0                  \\
\hline
\hline
          Silicon & This work                        & $42.80 \pm 0.03$   &  $154.93 \pm 0.04$      &  $69.42 \pm 0.02$    \\
                  & Experiment~\cite{Hall1967PR}     &  51              &  166                  &  64                 \\
                  & Experiment~\cite{cSiExp08}           &  52.2             &  171.5                &  67.1               \\
\hline
\multicolumn{5}{c}{} \\
\multicolumn{5}{c}{Amorphous samples} \\
\hline
System & Source &  $G$ [GPa]  & $C_{11}$ [GPa]  & $C_{12}$ [GPa] \\
\hline
\hline
\hline
          Silicon    & This work                        & $28.1 \pm 0.2$   &  $124.1 \pm 0.1$     &  $66.5 \pm 0.1$    \\
                     & Previous sim.~\cite{Hobbs04PRB}  &  25.2            &  131.0               &  81.4              \\
                     & Experiment~\cite{cSiExp1996}     &  48.8            &  156                 &  58.4              \\
\hline
\hline
           PMMA      & This work                        & $1.85 \pm 0.07$   &  $6.37 \pm 0.06$     &  $3.15 \pm 0.07$        \\
                     & Experiment~\cite{Cahill16Mac}     &  $2.0 \pm 0.1$  &  $9.6 \pm 0.9$     &      --               \\
\hline
\hline
          Cellulose  & This work                        & $2.71 \pm 0.03$   &  $14.84 \pm 0.02$     &  $8.62 \pm 0.03$        \\
\hline
\hline
 Cellulose Acetate   & This work                        & $2.09 \pm 0.01$   &  $10.85 \pm 0.02$     &  $6.59 \pm 0.01$        \\
\hline
\end{tabular}  \label{tab:elastic}
\end{center}
\end{table*}

A summary of the averaged $C_{ij}$ values, along with available literature data for comparison, 
is provided in Supplementary Table~\ref{tab:elastic}. 
Overall, the results exhibit good agreement with previously published values, 
thereby supporting the consistency and reliability of our computational approach.
In the case of silicon, our estimates align much better with previous simulations~\cite{Hobbs04PRB} than with experimental results~\cite{cSiExp1996}, 
where discrepancies are particularly large for the shear modulus of the amorphous solid.
We believe that this is mostly caused by deficiencies of the potential rather than by our comparatively large quench rates, which produce 
less stable structures than slower cooling protocols. 
We come to this conclusion, because changing the quench rate by an order of magnitude barely changes $G$, as shown in the Supplementary 
Section~S1 and Supplementary Figs. S1 and S2. 
Moreover, estimates of $C_{11}$ are already off by 10~GPa for the crystal, which has a well-defined reference structure. 
Thus, force-field induced errors of order 10~GPa in all elastic tensor elements must be expected.
However, what matters most for our work, in which we aim to establish a lean method for the computation of elastic constants, 
is that previous simulations~\cite{Hobbs04PRB} using similar potentials find similar values as we do.
Analyzing the shortcoming of the potential for a covalently bonded solid and suggesting how to mitigate deficiencies is beyond the scope of this work. 
Since non-covalent interaction, which primarily dictate the elastic tensor coefficients in polymeric systems, have a much 
reduced many-body character compared to interactions in covalently bonded solids, we expect our results on polymers to better compare to real systems than for silicon.

Notably, the values reported for cellulose and cellulose acetate are predictive in nature, as-- to the best of our knowledge-- 
no corresponding $C_{ij}$ data are currently available in the simulation or experimental literature. 
These predictions provide a valuable reference point for future studies. Furthermore, the computed 
$C_{ij}$ values for cellulose derivatives across a range of temperatures have already been successfully 
employed to estimate thermal transport coefficients $\kappa$ in related work~\cite{MukherjiMac26}, which was the
main motivation behind the development of the noise-cancellation approach proposed herein.
Lastly, we emphasize that kinetic-energy contribution to the computed $C_{ij}$ values are negligible in all cases, 
remaining below 0.1\%, and therefore do not influence the reported results.

The optimal approach will certainly be system--specific.
Metamaterials with extended domains of soft and stiff regions, which may allow liquid intrusions, 
are likely to require longer simulations for each thermostat time constant, particular care when applying the strain, 
and additional measures to make the noise--cancellation technique sufficiently efficient for meaningful results.
Discussing such complex systems in detail is beyond the scope of this work.
Nevertheless, it is clear that achieving the optimum balance between fast relaxation of the first moment of stress 
estimators and keeping stochastic errors small will be extremely challenging, since the first goal benefits from small 
damping while the second benefits from large damping. 

\section{Conclusions}
\label{sec:conclusions}

In this work, we have developed and applied a practical and generalizable method for accurately computing elastic constants 
at finite temperature, addressing common challenges associated with thermal noise and anharmonicity in molecular simulations. 
By extending a noise--cancellation approach previously applied to piezoelectric coefficients~\cite{Herzbach06cpc}, 
we have introduced a protocol that relies on computing stress differences between strained and unstrained (or oppositely strained) configurations, 
all evolved under identical thermostatting conditions. This setup effectively cancels out the dominant contributions of thermal 
fluctuations, allowing for low--variance estimates of elastic constants.
Importantly, our findings highlight a key advantage of this approach: it does not rely on fitting stress--strain curves or on time--averaged 
fluctuations of stress or strain, both of which are commonly hampered by noise and slow convergence at finite temperature.
This is particularly relevant for modern materials science, where increasingly complex materials -- ranging from amorphous semiconductors 
to bio--derived polymers -- require reliable mechanical characterization at the nanoscale, 
i.e., over volumes large enough to define local elastic properties yet small enough to capture heterogeneity in structure.

We validated the method across a diverse set of material systems, beginning with crystalline argon as a simple benchmark 
and progressing through ordered and amorphous silicon to structurally complex amorphous polymers such as poly(methyl methacrylate) and cellulose derivatives. 
In each case, our results show rapid and stable convergence of the elastic constants, with reasonable agreement with available reference data. 
For the polymeric systems, especially cellulose and cellulose acetate, our simulations provide the first calculation of the full elastic 
tensor at finite temperature, offering valuable new data for the materials modeling community.
We have further applied this method to determine the temperature--dependent elastic constants of a cellulose derivative and their direct 
implications for thermal transport~\cite{MukherjiMac26}, as well as the elastic constants of experimentally 
relevant commodity polymers under high pressures~\cite{Mukherji26JCP} and polyelectrolyte melts~\cite{Mukherji26Sub}.
While the current work utilizes a simple forward-difference scheme to demonstrate the robust performance of the noise 
cancellation technique even under non-ideal conditions, moving to a central-difference scheme would offer an immediate path to further increasing numerical stability. 
Such an approach would likely simplify the optimization of strain magnitudes by extending the range over which anharmonic effects remain negligible.

The method proposed in this work targets exclusively the (quasi--)static elastic constants.
The noise--cancellation technique still bears the possibility to deduce viscoelastic response functions.
To this end, a pair of constant--stress simulations would have to be performed, one under zero stress and another one under a constant perturbative stress. 
The (expectation value of the) difference in the strain between these two runs, divided by the stress increment, allows the time--dependent 
and thus frequency--dependent elastic modulus to be deduced.
The signal--to--noise ratio in this approach will certainly be much improved compared to a direct measurement 
without noise cancellation.
Nonetheless, decorrelation of the signal, which is expected to set in once the long--wavelength modes start relaxing, 
may require special measures to ensure meaningful results at medium and large time scales. 

Beyond its immediate application to elasticity, the methodology introduced here has potential extensions to other 
finite--temperature response properties, particularly the specific heat, whose calculation relies on differences analogous to those used for elastic constants. 
Furthermore, its simplicity and compatibility with standard molecular dynamics frameworks make it easily adaptable to a wide range of computational workflows.
We expect this approach to enable more accurate mechanical modeling across disciplines — including condensed matter physics, 
polymer science, and biomaterials engineering -- and to provide a foundation for future developments in finite--temperature materials simulations.\\

\noindent{\bf Acknowledgment:} 
D.M. thanks Tiago Espinosa de Oliveira for a collaboration~\cite{MukherjiMac26}, which provided
key systems investigated in this work.
Financial support for D.M. and M.M. was provided by the Bundesministerium f\"ur Technologie, Forschung und Raumfahrt (BMTRF) 
within the project 16ME0658K MExMeMo and European Union– NextGenerationEU. 
The authors gratefully acknowledge the Gauss Centre for Supercomputing e.V. (www.gauss--centre.eu) 
for providing computing time through the John von Neumann Institute for Computing (NIC) on the 
GCS Supercomputer JUWELS at J\"ulich Supercomputing Centre (JSC) where the polymer simulations were performed.
M.H.M. acknowledges helpful discussions with Sergey Suhomlinov and Lars Pastewka. 

\bibliographystyle{ieeetr}
\bibliography{MukherjiarXiv}

\end{document}